\documentclass[twocolumn,aps,showpacs,preprintnumbers,amsmath,amssymb, superscriptaddress]{revtex4-1}
\pdfoutput=1

\usepackage{bm}
\usepackage{graphicx}
\usepackage{color}

\begin{document}

\title{Three-Dimensional Nanoporous Graphene Substrate for Surface-Enhanced Raman Scattering}

\author{Zhiqiang Tu}
\affiliation{State Key Laboratory of Heavy oil Processing, China University of Petroleum, Beijing, 102249, China}
\author{Shangfei Wu}
\affiliation{Beijing National Laboratory for Condensed Matter Physics, and Institute of Physics, Chinese Academy of Sciences, Beijing 100190, China}
\author{Fan Yang}
\affiliation{State Key Laboratory of Heavy oil Processing, China University of Petroleum, Beijing, 102249, China}
\author{Yongfeng Li}\email{yfli@cup.edu.cn}
\affiliation{State Key Laboratory of Heavy oil Processing, China University of Petroleum, Beijing, 102249, China}
\author{Liqiang Zhang}
\affiliation{State Key Laboratory of Heavy oil Processing, China University of Petroleum, Beijing, 102249, China}
\author{Hongwen Liu}\email{hliu@iphy.ac.cn}
\affiliation{Beijing National Laboratory for Condensed Matter Physics, and Institute of Physics, Chinese Academy of Sciences, Beijing 100190, China}
\author{Hong Ding}
\affiliation{Beijing National Laboratory for Condensed Matter Physics, and Institute of Physics, Chinese Academy of Sciences, Beijing 100190, China}
\affiliation{Collaborative Innovation Center of Quantum Matter, Beijing, China}
\author{Pierre Richard}
\affiliation{Beijing National Laboratory for Condensed Matter Physics, and Institute of Physics, Chinese Academy of Sciences, Beijing 100190, China}
\affiliation{Collaborative Innovation Center of Quantum Matter, Beijing, China}

\date{\today}

\begin{abstract}
We synthesized three-dimensional nanoporous graphene films by a chemical vapor deposition method with nanoporous copper as a catalytic substrate. The resulting nanoporous graphene has the same average pore size as the underlying copper substrate. Our surface-enhanced Raman scattering (SERS) investigation indicates that the nanoporosity of graphene significantly improves the SERS efficiency of graphene as a substrate as compared to planar graphene substrates.  
\end{abstract}

\pacs{74.70.Xa, 74.25.Jb, 79.60.-i}


\maketitle

Surface-enhanced Raman scattering (SERS), which consists in the enhancement of inelastic light scattering of molecules adsorbed on a surface, has attracted much interest as an analytical method owing to its ability to provide simultaneously structural and quantitative information for analytes. Most often, this effect is generated by the strong local electromagnetic confinement resulting from rough metallic surfaces or nanostructures such as colloidal silver or gold nanoparticles \cite{PC_Lee_JPC86,Freeman_Science267} and dealloyed, nanoporous metallic films \cite{LY_Chen_AFM19,HW_Liu_SciRep1}. Compared with metal substrates, graphene has many advantages such as lower cost and better biocompatibility. In principle though, graphene as a SERS substrate does not support the electromagnetic mechanism (EM) because the surface plasmons on graphene are in the terahertz range rather than in the visible range \cite{Rana_IEEE2008}. Instead, SERS arises from a charge transfer between the adsorbed molecules and a substrate with proper chemical affinity with these molecules. Indeed, SERS has been reported recently for two-dimensional single-layer graphene \cite{X_Ling_NanoLett10,X_Ling_Small6,X_Ling_JPCC116,X_Ling_JPCC117}. Interestingly, three-dimensional (3D) architectures of graphene have been gaining much attention recently \cite{Nardecchia_CSR42}, and continuous 3D-foam-like graphene with several hundred micrometers pore size have been synthesized by the metal-catalyzed chemical vapor deposition (CVD) method using Ni substrate \cite{ZP_Chen_Nmat10, N_Li_PNAS109}. Considering the positive role of nanoporosity on SERS and motivated by the progresses on graphene, here we present a study of SERS enhancement for nanoporous graphene (NG) with a 3D network of nanometer pore sizes. Our results indicate a significant SERS enhancement as compared to planar graphene and nanoporous copper (NC).

The 3D NG used in this study has been synthesized by a CVD method with NC as a catalytic substrate. In order to synthesize 3D NG, NC with different size distributions were first fabricated by dealloying 0.14 mm-thick Cu$_{30}$Mn$_{70}$ alloys, synthesized by a melt spinning method \cite{Weissmuller_MRS_Bull34, Hayes_JMatRes21, Erlebacher_Nature410}. The pore size of NC could be adjusted by changing the dealloying time of the erosion aqueous solution consisting of (NH$_4$)$_2$SO$_4\cdot$H$_2$O (1 M$\cdot$L$^{-1}$) and MnSO$_4\cdot$H$_2$O (0.01 M$\cdot$L$^{-1}$). Then, NC substrates were used to synthesize NGs by a CVD process using a mixture of gases with a composition of CH$_4$: Ar: H$_2 = 8:200:20$ sccm. Owing to the high activity of NC, the growth temperature was lowered to 600-900 $^{\circ}$C. Finally, to investigate the SERS effect from 3D NG itself without the influence of the NC substrate, the 3D NG was released by etching graphene/NC stacks in aqueous FeCl$_3$ solution and then rinsed in deionized water and transferred to a Si substrate.

\begin{figure}[!t]
\begin{center}
\includegraphics[width=3.4in]{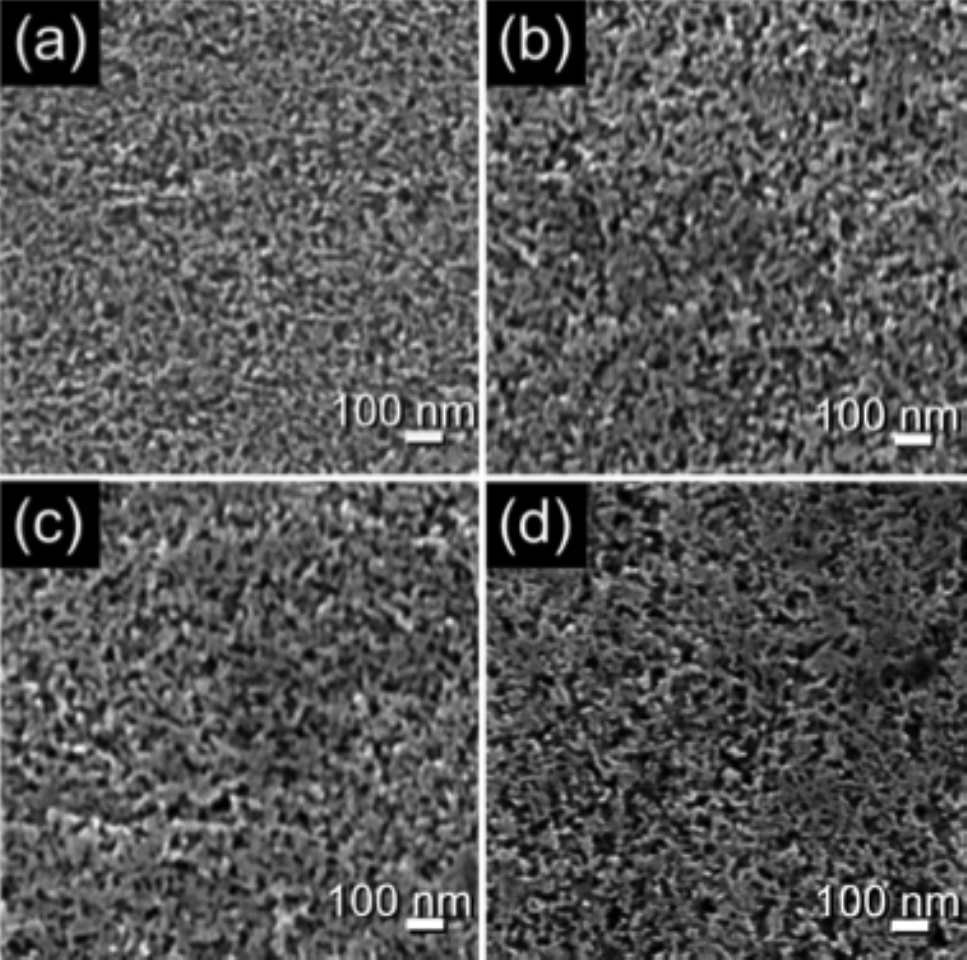}
\end{center}
\caption{\label{N_copper}(Color online). SEM images of nanoporous coppers with different pore sizes: a) 23 nm, b) 28 nm, c) 33 nm, and d) 40 nm.}
\end{figure}

\begin{figure}[!t]
\begin{center}
\includegraphics[width=3.4in]{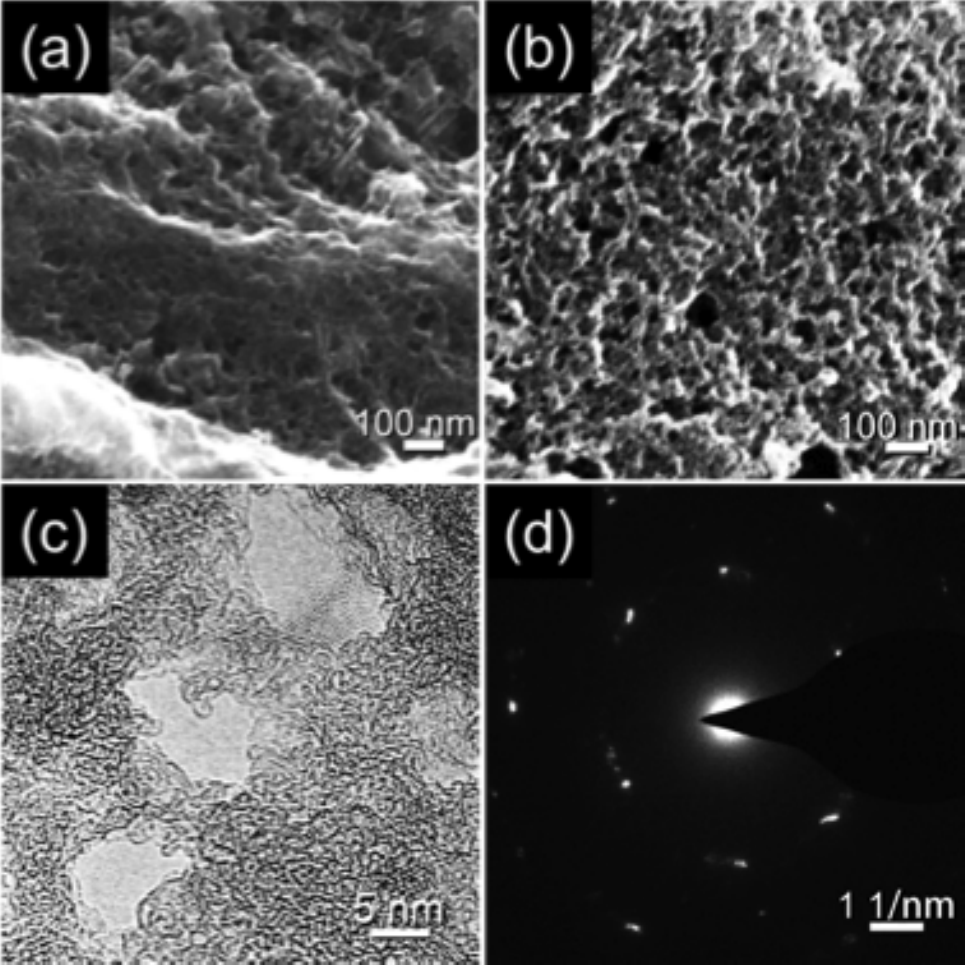}
\end{center}
\caption{\label{NG}(Color online). SEM images of NG with a) a side view, and b) a top view. c) A typical TEM image, and d) a SAED image of NG.}
\end{figure}

Figure \ref{N_copper} shows the scanning electron microscopy (SEM) images of NC with average pore sizes ranging from 23 to 40 nm. The SEM image of sectional NC provides more clear nanoporosity morphology, as shown in Fig. S1. Figure \ref{NG}a shows the cross-sectional SEM image of NG with clear porous structures. The nanopore size of NG (Fig. \ref{NG}b) is in accordance with the front view of the morphology of NC. As shown in Fig. \ref{NG}c and Fig. S2, our transmission electron microscopy (TEM) characterization of the NG films indicates that they form a continuous 3D porous structure consisting of one to several layers of graphene. The selected area electron diffraction image displayed in Fig. \ref{NG}d exhibits hexagonal spots in one circle, implying that the NG has a single crystal structure. From these experiments we conclude that the NG films grown on the NC surfaces have similar nanoporous structure as the underlying NC templates.

We further study the NG films using a Horiba JY-T64000 Raman spectrometer. Spectra were recorded using 514.5 nm laser light and a 100$\times$ objective illuminating a 0.8 $\mu$m$^{2}$ spot size. Typical monolayer graphene or few-layers graphene Raman spectra above 1000 cm$^{-1}$ are characterized by a strong G peak around 1580 cm$^{-1}$, which corresponds to the first-order scattering of E$_{2g}$ phonon of $sp^2$ carbon atoms. In the presence of defects, a peak called D band ($\kappa$-point phonons of A$_{1g}$ symmetry) is also observed around 1350 cm$^{-1}$, as well as a 2D band corresponding to second-order scattering involving defects \cite{FerrariPRL97}.

\begin{figure*}[!t]
\begin{center}
\includegraphics[width=7in]{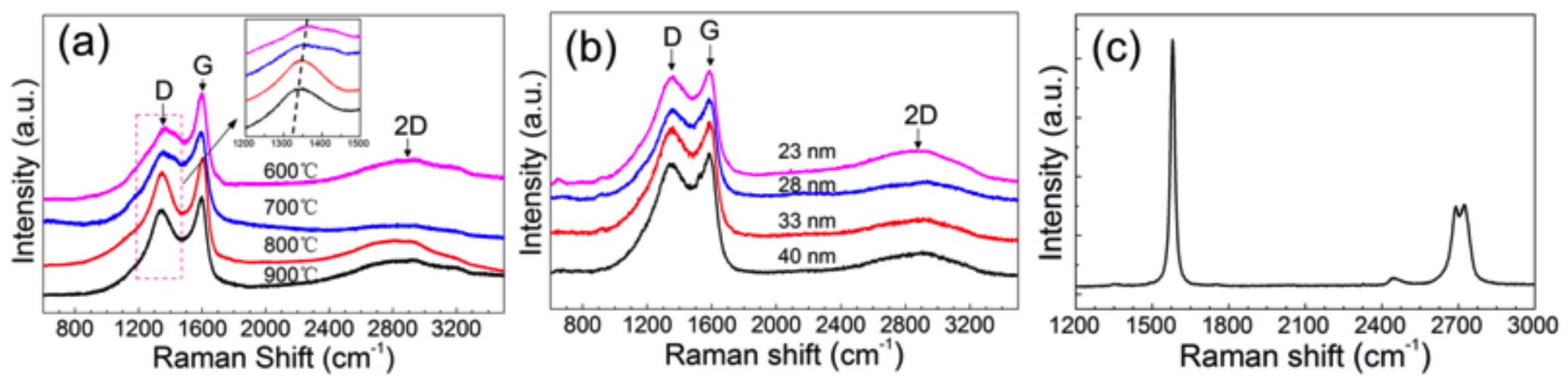}
\end{center}
\caption{\label{Raman_NG}(Color online). a) Raman spectra of 3D NG samples synthesized at different temperatures on a 33 nm NC substrate. b) Raman spectra of NG samples with different nanopore sizes prepared at 800 $^{\circ}$C. c) A Raman spectrum of microporous graphene. The laser power is 30 mw for a)-c).}
\end{figure*}

In Fig. \ref{Raman_NG}a we show the Raman spectra of NG films synthesized at different temperatures via NC substrates with pore size of 33 nm. A slight increase in the G peak position is observed with the synthesis temperature increasing, in contrast to a 1364 to 1347 cm$^{-1}$ downshift of the D peak, which shows high intensity attributed to defects \cite{FerrariPRL97}. In addition, a slight shoulder appearing near the D band is observed in the NG samples synthesized at 600 and 700 $^{\circ}$C, which implies enhanced disorder \cite{Tuinstra_JCP53}.

Raman spectra of NG samples synthesized on NC substrates with different pore sizes at the optimized synthesis temperature of 800 $^{\circ}$C are shown in Fig. \ref{Raman_NG}b. Strong and broadened D and G modes are observed and a broad 2D mode is also found. These peaks broadness and relative intensities are notably different from the typical spectra of monolayer or few-layer two-dimensional graphene, which can be understood in terms of high disorder induced from edges and clusters with different dimensions in NG films \cite{Ryu_ACSnano5, FerrariPRB61}. Compared with planar graphene film, the frequency of the G peak shows some shift towards high wavenumber, and is always observed around 1590 cm$^{-1}$ in our samples. These results are possibly due to a decrease in the two-dimensional crystal size and an increase in the number of edges on the surface of NG \cite{FerrariPRL97, Tuinstra_JCP53}. For comparison, the Raman spectrum of a microporous graphene grown on a microporous copper substrate, displayed in Fig. \ref{Raman_NG}c has a much sharper and narrower 2D peak. Since the intensity ratio of the D and G peaks, $I_D/I_G$, is related to the crystallite size, Eq. \eqref{crystal_size} has been developed to calculate the crystal size of nanographene or nanographite, 

\begin{equation}
\label{crystal_size}
L_a (nm)=(2.4\times10^{-10})\lambda^4(I_D/I_G)^{-1}
\end{equation}

\noindent where $\lambda$ is the laser wavelength in nanometers \cite{Cancado_APL88}. According to this equation, the crystal size of NG falls in the 3-9 nm range. In addition, though the 2D band shape is not repeatable for the different measured regions, it is common to observe a broad 2D band between 2400 and 3300 cm$^{-1}$.

\begin{figure*}[!t]
\begin{center}
\includegraphics[width=7in]{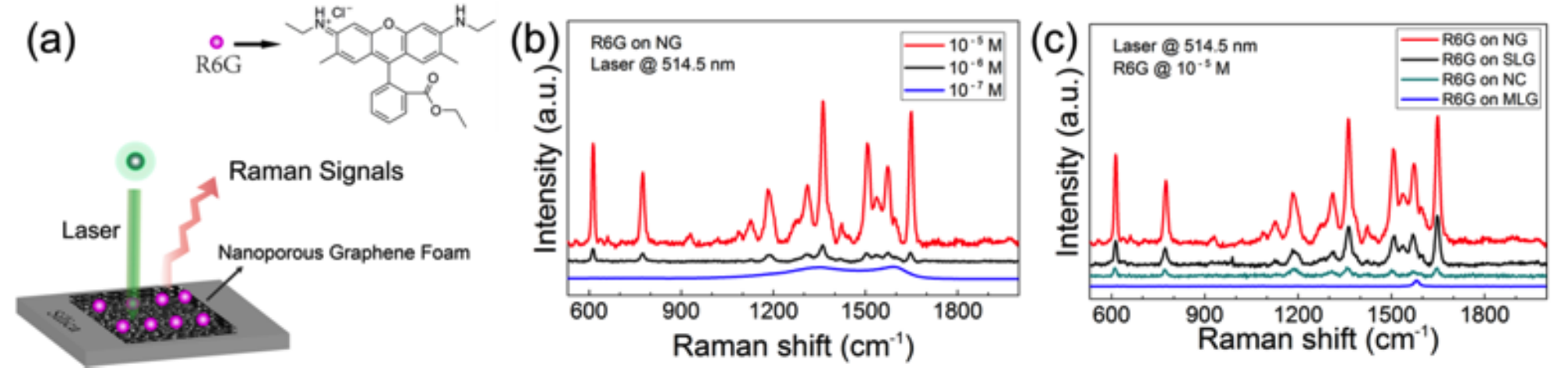}
\end{center}
\caption{\label{Fig_SERS}(Color online). a) Scheme of Raman detection of R6G on a plain 3D NG substrate. Raman signals of R6G collected b) from different concentrations, and c) on different substrates. Same average pore size of 33 nm has been used for all the nanoporous films. We used laser power of 30 mw for b) and c). The fluorescence background of R6G in b) and c) has been removed from the spectra.}
\end{figure*}

We now discuss the SERS properties of NG using rhodamine 6G (R6G) as probed molecules. In order to exclude the plasmonic influence from the NC substrates, the NG films have been transferred from the NC substrates onto SiO2/Si substrates. Furthermore, different concentrations of R6G were deposited on the 3D NG substrates by soaking them in R6G solutions for 30 minutes. After soaking, the samples were dried naturally. The setup for Raman detection of R6G on a plain 3D NG substrate is schematically drawn in Fig. \ref{Fig_SERS}a.

\begin{table}
\caption{\label{Table_Raman}Assignments of Raman peaks of 10-5 M R6G on NG and NC excited at 514.5 nm.}
\begin{ruledtabular}
\begin{tabular}{ccc}
On NG (cm$^{-1}$)&On NC (cm$^{-1}$)&Assignment\\
\hline
613&610&C-C-C ring in-plane bending\\
775&772&C-H out-of-plane bending\\
1182&1182&aromatic C-C stretching\\
1277&&C-O-C stretching\\
1310&1309&C-O-C stretching\\
1361&1359&aromatic C-C stretching\\
1507&1500&aromatic C-C stretching\\
1573&1567&aromatic C-C stretching\\
1648&1648&aromatic C-C stretching\\
\end{tabular}
\end{ruledtabular}
\end{table}

We tested the Raman signals of 10$^{-6}$ M R6G with 488, 514.5, and 632 nm excitations (Fig. S3). The results show the strongest signal for 514.5 nm laser light, which is consistent with the resonant wavelength of the R6G molecules \cite{Guthmuller_JPCA112}. To estimate the SERS capability of a plain NG substrate, we decreased the concentration of the R6G solution from $1\times 10^{-5}$ M down to $1\times 10^{-7}$ M and recorded spectra in the same conditions of incident light power and acquisition time. The corresponding Raman spectra shown in Fig. \ref{Fig_SERS}b are the average results, which are obtained from a $\sim 1$ $\mu$m$^2$ region of the sample equal to the laser spot size. For the highest R6G concentration of $1\times 10^{-5}$ M, the Raman spectrum exhibits several sharp fingerprints in the measured range that are characteristic of the R6G molecule. The energy position of the most prominent ones and their assignments \cite{Weiss_JPCB105, Hildebrandt_JPC88, WL_Zhai_Nanoscale4} are given in Table \ref{Table_Raman}. Apart for a lower intensity, the spectrum recorded with the film soaked in the $1\times 10^{-6}$ M R6G solution is similar, with the same R6G modes observed. We note that the relative intensities of the peaks at 1361, 1507, 1573 and 1648 cm$^{-1}$ vary from the 10$^{-5}$ M to the 10$^{-6}$ M soaked samples, which is a typical signature in SERS for a small number of molecules (usually for 10$^{-6}$ M and lower concentrations), probably due to the inhomogeneous interactions between the substrate and the molecules \cite{Kneipp_PRL78}. In contrast, no Raman R6G peak is detected when using the $1\times 10^{-7}$ M R6G solution, which only shows the broad D and G modes of the NG substrate. However, we caution that stronger local SERS enhancement may occur at few hot spots in the curled NG film \cite{HW_Liu_SciRep1}. We note that the spectra corresponding to higher concentrations also show a broad background around the energies of the D and G modes, which we also attribute to the substrate.

In order to compare the SERS efficiency of different substrates, we show in Fig. \ref{Fig_SERS}c the Raman spectra of R6G with 10$^{-5}$ M obtained on NG (33 nm average pore size), NC (33 nm average), multi-layer graphene (MLG, 10-20 layers, CVD method) and single-layer graphene (SLG) substrates. Showing only a sharp G peak at 1581 cm$^{-1}$ from the substrate and no signature of R6G, MLG has the lowest efficiency of all these materials. Improvement is obtained using NC. Although the R6G peak positions are consistent with those obtained for the NG substrate though, as reported in Table \ref{Table_Raman}, the Raman intensity remains weak and the peaks are broad, which is particularly noticeable for the mode at 610 cm$^{-1}$. The best results, with a large number of sharp peaks, are obtained with the NG and SLG substrates. However, Fig. \ref{Fig_SERS}c clearly indicates that the SERS intensity is stronger by a factor of about $3\sim 4$ when using the former substrate instead of CVD-grown SLG, which is characterized by TEM and Raman spectroscopy characterizations \cite{ZQ_Tu_Carbon73}.

We quantitatively calculate the enhancement factor (EF) of the nanoporous graphene film for the Raman peaks at $\sim$1650 and $\sim$610 cm$^{-1}$ of R6G on a nanoporous graphene film (the detailed method is described in the supporting information). The EFs are estimated to be about 47. Our observation of SERS enhancement in the sequence is NG $>$ SLG $>>$ NC $>>$ MLG. The origin of this enhancement remains unclear. It has been reported that the excitation of intrinsic graphene plasmons are tunable and strengthened by natural nanoscale inhomogeneities, such as wrinkles \cite{Grigorenko_Nphoton6, Crassee_NanoLett12}. Most likely, the better SERS efficiency of NG over SLG suggests that enhancement of the local electromagnetic field due to porosity and surface roughness plays an important role in tuning the plasmonic properties of these materials. The big curvature of pores breaks the perfect structure of graphene and cavities that are formed in the nanopores, which may introduce and strengthen an extra local electric field on molecules \cite{Maier2007}. Tuning chemical enhancement in SERS in this way has been reported on graphene oxide by introducing oxygen defects \cite{X_Yu_ACS_Nano5}. We carefully consider the possibility that the pores of the nanoporous substrates increase the amount of R6G molecules adsorbed on the surface. It improves the limited efficiency of molecule detection in nano-devices, but this mechanism cannot be viewed as intrinsic SERS enhancement. Finally, the strong SERS enhancement due to the chemical enhancement mechanism, \emph{i.e.}, the $\pi-\pi$ interactions between the R6G molecules and the carbon-based substrates cannot be neglected, SLG being more efficient than NC. We emphasize that at the current stage the enhancement factor of graphene should not be expected as comparable with that of some metallic films like gold because graphite itself is an inactive plasmonic material.

3D-foam-like graphene films with tunable nanopore sizes have been synthesized by a CVD method with NC as substrates, and show that the resulting nanoporous graphene has the same average pore size as the underlying copper substrate. The Raman enhancement effect has been investigated by using R6G molecules. Our results indicate that NG has a great potential use for SERS substrate that is better than NC, MLG and even SLG. Considering the special spatial structure of NG, chemical enhancement and electromagnetic mechanisms, as well as molecule trapping in pores can be attributed to the effective SERS effect. This new kind 3D graphene network is expected to broaden the study in nano-electronic devices, especially in microanalysis.

This work was supported by the National Natural Science Foundation of China (Nos. 21322609, 51071175, 21106184 and 11274362), the Ministry of Science and Technology of China (No. 2011CBA001000) and Thousand Talents Program.

\bibliography{biblio_HW}

\end{document}